\documentclass[aps,prl,notitlepage,twocolumn,superscriptaddress,longbibliography,nofootinbib]{revtex4-1}
\pdfoutput=1
\usepackage[utf8]{inputenc}
\usepackage{amsmath,amsthm,amssymb,amsfonts}
\usepackage{scalerel}
\usepackage{mathtools}
\usepackage{graphicx}
\usepackage{subfigure}
\usepackage[colorlinks = true,
            linkcolor = blue,
            urlcolor  = blue,
            citecolor = blue,
            anchorcolor = blue]{hyperref}
\usepackage{mathrsfs}
\usepackage{bbold}
\usepackage[]{units}
\usepackage{bm}
\usepackage{braket}
\usepackage{color}
\usepackage{makecell}
\usepackage{enumitem}
\usepackage{upgreek}
\usepackage{blindtext}
\usepackage{graphics}
\usepackage{verbatim} 
\usepackage{algorithm}
\usepackage[normalem]{ulem}
\usepackage[noend]{algpseudocode}

\usepackage[dvipsnames]{xcolor}
\usepackage{bbm}
\usepackage[bb=boondox]{mathalfa}
\usepackage{array}
\usepackage{multirow}
\usepackage{tabularx}
\usepackage{float}
\usepackage{graphicx}
\usepackage{dcolumn}
\usepackage{bm}
\usepackage{amsmath,amsfonts,amssymb}
\usepackage{slashed}
\usepackage{braket,xcolor}
\usepackage{verbatim}
\usepackage{multirow}
\usepackage{amsfonts, mathtools,resizegather}
\usepackage[export]{adjustbox}

\begin{document}

\title{{Krylov fractality and complexity in generic random matrix ensembles}}
\author{Budhaditya Bhattacharjee\,\,\href{https://orcid.org/0000-0003-1982-1346}
{\includegraphics[scale=0.05]{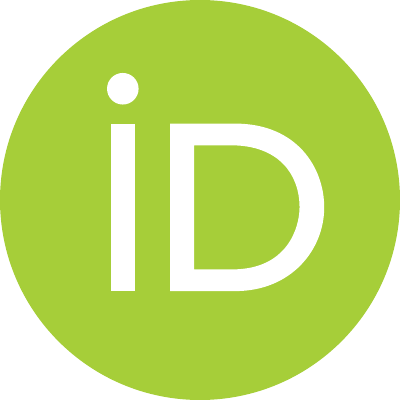}}\,}
\email{budhadityab@ibs.re.kr}
\affiliation{Center for Theoretical Physics of Complex Systems,\\ Institute for Basic Science (IBS),  Daejeon 34126, Republic of Korea}
\author{Pratik Nandy\,\,\href{https://orcid.org/0000-0001-5383-2458}
{\includegraphics[scale=0.05]{orcidid.pdf}}\,}
\email{pratik@yukawa.kyoto-u.ac.jp}
\affiliation{Center for Gravitational Physics and Quantum Information, Yukawa Institute for Theoretical Physics,\\ Kyoto University, Kitashirakawa Oiwakecho, Sakyo-ku, Kyoto 606-8502, Japan}
\affiliation{RIKEN Interdisciplinary Theoretical and Mathematical Sciences Program (iTHEMS),
Wako, Saitama 351-0198, Japan}

\begin{abstract}

Krylov space methods provide an efficient framework for analyzing the static and dynamical aspects of quantum systems, with tridiagonal matrices playing a key role. Despite their importance, the behavior of such matrices from chaotic to integrable states, transitioning through an intermediate phase, remains unexplored. We aim to fill this gap by considering the properties of the tridiagonal matrix elements and the associated basis vectors for appropriate random matrix ensembles. We utilize the Rosenzweig-Porter model as our primary example, which hosts a fractal phase in addition to the ergodic and localized regimes. We discuss the characteristics of the matrix elements and basis vectors across these three (ergodic, fractal, and localized) regimes and introduce Krylov space tools to identify the transition points. The exact expressions of the Lanczos coefficients of such tridiagonal matrices are provided in terms of $q$-logarithmic function across the full parameter regime. The numerical results are corroborated with analytical reasoning for certain features of the Krylov spectra. Additionally, we investigate the Krylov state (spread) complexity within these regimes, showcasing the efficacy of our methods in pinpointing these transitions.

\end{abstract}

~~~~~~~~~~~~~~~~~~~~~~~ RIKEN-iTHEMS-Report-24

\maketitle

\emph{Introduction:} A large extent of our understanding of quantum systems stems from the study of random matrices that closely model the generators of the dynamics of the system \cite{mehta1991random}. Pioneering conjectures by Bohigas-Giannoni-Schmidt \cite{BGS} and Berry-Tabor \cite{Berry_Tabor_conj} have laid the groundwork for understanding level statistics \cite{Wigner1, Dyson1962a}, while random matrix theories have emerged as potent models for elucidating black hole physics \cite{Cotler:2016fpe}. Moreover, the transitions between chaotic and integrable phases in quantum systems have been a focal point of research, with the eigenspectrum of the Hamiltonian providing valuable insights into these transitions \cite{olshanii2012exactly, Kravtsov_2015, das2023absence, das2023robust}.

This Letter addresses phase transitions within a prototypical random matrix ensemble, employing Krylov space techniques \cite{Parker:2018yvk, Balasubramanian:2022tpr, Nandy:2024htc}. These techniques have gained prominence due to their simplicity and wide applicability; see \cite{Barbon:2019wsy, Dymarsky:2019elm, Rabinovici:2020ryf, Rabinovici:2021qqt, Dymarsky:2021bjq, Caputa:2021sib, Bhattacharjee:2022vlt, Hornedal:2022pkc, Bhattacharya:2022gbz,  Liu:2022god, Bhattacharjee:2022qjw, Bhattacharjee:2022lzy, Avdoshkin:2022xuw, Camargo:2022rnt, Erdmenger:2023wjg} for the initial explorations and \cite{Nandy:2024htc} for a detailed review and the references therein. By transforming the Hamiltonian into a tridiagonal form, one analyzes its elements, known as Lanczos coefficients, which comprise the Krylov spectrum \cite{Parker:2018yvk, Balasubramanian:2022tpr}. Such techniques have been employed in a few particular cases in different phases \cite{ scialchi2024integrability, Alishahiha:2024rwm, Menzler:2024ifs, Cohen:2024ngg}. However, the behavior across the transition regime remains unexplored. We address such transition through Krylov space tools in the Rosenzweig-Porter model \cite{PhysRev.120.1698, PhysRev.120.1698, altland1997perturbation, khaymovich2021dynamical, pino2019ergodic} which is particularly intriguing as it exhibits a spectral fractal phase, bridging the gap between ergodic and localized regimes. We propose approximate \emph{Ans\"atze} that characterize the statistical behavior of the Lanczos coefficients during the transitional phases. The \emph{static} inverse participation ratio (IPR) \cite{IPRref} of Krylov vectors is introduced to assess the degree of localization, uncovering the IPR's ability to differentiate among the three distinct phases. We also investigate the complexity associated with the spread of a Gibbs state through these phases, identifying specific features that sharply delineate phase boundaries.

\emph{Rosenzweig-Porter model:} The Rosenzweig-Porter (RP) model \cite{PhysRev.120.1698} is given by the $N \times N$ Hamiltonian \cite{replicaScipost}
\begin{align}
    H = A + \frac{1}{N^{\gamma/2}} B\,, \label{RPmodel}
\end{align}
where $A$ is a diagonal matrix, drawn from a normal distribution with zero mean and unit variance, i.e., $\braket{a_{ii}}^2= 1$, and $B$ is a matrix from the Gaussian orthogonal ensemble (GOE), drawn from a normal distribution with zero mean (for both diagonal and off-diagonal elements) and variance $\braket{b_{ii}}^2= 1$ and $\braket{b_{ij}}^2 = 1/2$ for $i \neq j$, respectively. Here $\gamma \geq 0$ is a parameter that effectively controls the dominant behavior of $A$ and $B$. It gives rise to the corresponding delocalized or ergodic ($\gamma \in [0,1]$), fractal ($\gamma \in [1,2]$), and localized ($\gamma \in [2,\infty)$) phases, which leave imprints on the eigenspectrum of $H$. In the large-$N$ limit, one can obtain the $\braket{r}$-value statistics of the midspectrum \cite{Atas2013distribution}. Specifically, one obtains the Wigner-Dyson distribution for $\gamma = [0,2)$ while the Gaussian distribution appears in the $\gamma = (2,\infty)$ limit \cite{altland1997perturbation}; see Supplemental Material (SM1) \cite{supp} for further details. On the other hand, the fractal phase is characterized by the \emph{fractal dimension} $\mathcal{D}$, defined through the scaling of the IPR as $\mathrm{IPR} = \sum_n |\braket{n|\psi}|^{4} \sim N^{-\mathcal{D}}$ \cite{Buijsman:2021xbi}. Here $\ket{\psi}$ is a state in the eigenbasis and $\ket{n}$ is a computational basis element, and the overlap captures the spread of the eigenstate in the computational basis. The ergodic and localized regime correspond to constant fractal dimensions, i.e., $\mathcal{D} = 1$ and $\mathcal{D}=0$, respectively, while in the fractal phase, the fractal dimension linearly decreases as $\mathcal{D} = 2-\gamma$ \cite{Kravtsov_2015}.

\emph{Tridiagonalization and Krylov Spectrum: } To investigate the various phases of the RP model, we employ the Krylov spectrum of the Hamiltonian as our primary set of tools. The Hamiltonian is transformed into a tridiagonal matrix on the Krylov basis through the application of the Lanczos algorithm \cite{Lanczos1950AnIM, viswanath1994recursion}
\begin{equation}
\label{lanczosH}
     H \ket{K_n} = \mathsf{b}_{n} \ket{K_{n -1}} + \mathsf{a}_n \ket{K_n} + \mathsf{b}_{n + 1}\ket{K_{n +1}}\,.
\end{equation}
Here, the sets $\{\mathsf{a}_{n}, \mathsf{b}_{n}\}$ are referred to as the Lanczos coefficients, and the set $\{\ket{K_n}\}$ constitutes the Krylov basis, which together forms the Krylov spectrum \cite{Balasubramanian:2022tpr}; see \cite{Nandy:2024htc} for a detailed review. For numerical purposes, it is often useful to consider the Hessenberg decomposition \cite{Hessenberg}, which uses the Householder reflections \cite{householder1958} instead of the Lanczos algorithm. This produces identical results for Lanczos coefficients, subject to some redefinition of Krylov basis vectors. Curiously, the density of the states (DOS) can be well approximated from the statistics of the Lanczos coefficients $\mathsf{a}(x)$ and $\mathsf{b}(x)$ where $x = n/N$ in the large-$N$ limit as \cite{Balasubramanian:2022dnj, Balasubramanian:2023kwd}
\begin{align}
    \rho (E) = \int_{0}^1 dx \,\frac{\Theta(4 \mathsf{b}(x)^2 - (E - \mathsf{a}(x))^2)}{\pi \sqrt{4 \mathsf{b}(x)^2 - (E - \mathsf{a}(x))^2}}\,. \label{den0}
\end{align}
Here $\Theta(z)$ is the Heaviside theta function such that $\Theta(z) = 1$ for $z \geq 0$ and $\Theta(z) = 0$ for $z < 0$. Given the Lanczos coefficients, the Krylov wavefunctions $\uppsi_n (t)$ satisfy the following differential equation:
\begin{align}
\label{amplitudesH}
    i  \partial_t \uppsi_n (t) = \mathsf{b}_{n} \uppsi_{n-1}(t) + \mathsf{a}_{n}\uppsi_n(t) + \mathsf{b}_{n +1} \uppsi_{n +1}(t) \,.
\end{align}
which is the Schr\"odinger equation in the ``Krylov chain'' governed by the Hamiltonian \eqref{RPmodel}, and the time-evolved state is expressed in the Krylov basis $\ket{K_n}$ with the amplitude $\uppsi_n (t)$. The Krylov state (spread) complexity is defined by the average position on the Krylov chain \cite{Balasubramanian:2022tpr}
\begin{align}
    K_S(t) = \sum_n n \,|\uppsi_n (t)|^2\,.
\end{align}
The initial state plays an important role in governing the dynamics. In principle, one can start with any state, including the thermofield double (TFD) state or thermal Gibbs state, in which case the spread complexity shows distinct behavior: early quadratic growth followed by linear growth until it reaches a peak. The presence of the peak is crucial for chaotic systems and differs from integrable systems where such a peak is absent \cite{Balasubramanian:2022tpr, Erdmenger:2023wjg}. At late times, the spread complexity saturates to a plateau, depending on the dimension of the Hilbert space.

\begin{figure}[t]
\hspace*{-0.5cm}
\includegraphics[width=0.5\textwidth]{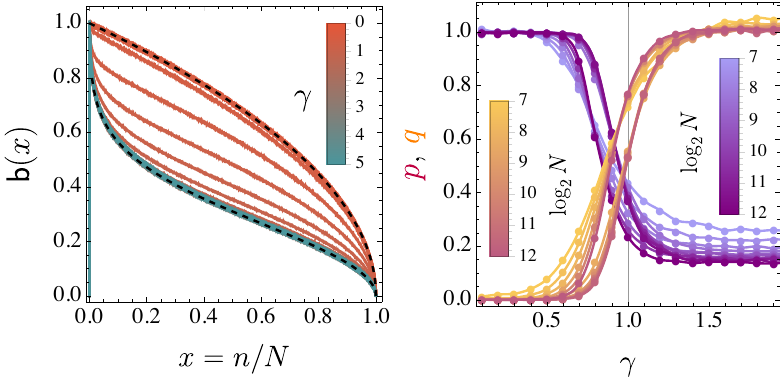}
\caption{(Left) The Lanczos coefficients for the RP model with $N  = 4096$ (with $100$ ensemble averages) as a function of $x = n / N$, where $n$ is the index of the Lanczos coefficient. The black dashed lines are the fitted results given by \eqref{bngoe} (top) and \eqref{bnpois} (bottom) at the two extremes. (Right) The behavior of the fitting parameters $u$ and $v$ as functions of $\gamma$ and $N$.} \label{fig:lanczosinitialfull}
\end{figure}

\emph{Statistics of Lanczos coefficients: } 
In a manner akin to the conventional eigenspectrum analysis, we focus on the behavior of the Lanczos coefficients for the Hamiltonian \eqref{RPmodel} throughout all phases, including the critical points of fractal and localization transitions. 
We choose the initial state as $(1,0,0, \cdots,0)^{\intercal}$. However, the choice of the initial state does not affect the \emph{averaged} or \emph{mean} of the Lanczos spectrum \emph{in the bulk}, given a sufficient number of ensemble averages of the Hamiltonian is taken. This is because the averaged bulk Lanczos coefficients depend solely on the density of states of the Hamiltonian. However, even after a large number of ensemble averages, the edge coefficients, i.e., $n \sim o(1)$ coefficients, may exhibit a slight dependence on the choice of the initial state in small system sizes. This effect diminishes as the system size increases.

Figure \ref{fig:lanczosinitialfull} shows the behavior of the \emph{rescaled} Lanczos coefficients in the RP model as a function of $\gamma$. As the largest eigenvalue scales with system size $N$, selecting larger system sizes will increase the largest eigenvalue, making the Lanczos coefficients dependent on $N$. For example, in the case of GOE, $b_n \sim \sqrt{N/2}$ with $n \sim o(1)$ if the variance of GOE matrices is taken as unity. Conversely, if the variances are taken as $2/N$, then $b_n \sim 1$ when $n \sim o(1)$, and does not scale with system size. To ensure this fact, we choose appropriate variances for the random matrices so that the Lanczos coefficients are expressed in a system size-independent manner.

We observe a transition in the behavior of the Lanczos coefficients as $\gamma$ transits from the ergodic to the localized phase. The behavior in the ergodic phase (where the matrix is full GOE) is known analytically \cite{zuker2001canonical, Balasubramanian:2022dnj} and given by
\begin{align}
    \mathsf{b}_{\gamma = 0} (x) \equiv \mathsf{b}_{\text{GOE}} (x) \propto \sqrt{1 - x}\,. \label{bngoe}
\end{align}
The symmetry of the energy in the DOS requires $\mathsf{a}(x) = 0$ upon the ensemble average of the Hamiltonian. In the localized phase $\gamma \gtrsim 2$, the nature of the function is previously unknown. In this Letter, we ascertain that such a form is given by
\begin{align}
    \mathsf{b}_{\gamma \gtrsim 2} (x) \equiv \mathsf{b}_{\text{P}} (x) = \xi \sqrt{d}\, \text{nib}(x,d)\,, \label{nib1}
\end{align}
where $d = \log_2 N$ and $N$ is the matrix dimension. Here ``$\text{nib}$'' stands for the inverse of the shifted binomial function $\text{bin}(x, d) = 2^{-d} \binom{d}{d(1/2 - x)}$ \cite{zuker2001canonical}. Such structure has been previously studied in the context of nuclear spectra \cite{molina2005spectral, zuker2001canonical}, yet remains largely unknown. The inverse binomial is ill defined at the origin, while away from $x \sim 0$, it can be approximated to a simpler form
\begin{align}
    \mathsf{b}_{\text{P}} (x) \sim \sqrt{-\frac{\xi^2}{2}\ln x}\, ~\Leftrightarrow~ \rho(E) = \frac{1}{\sqrt{2\pi} \xi} e^{-E^2/(2\xi^2)}\,, \label{bnpois}
\end{align}
with corresponding Gaussian DOS of zero mean and variance $\xi^2$, obtained through \eqref{den0}; see SM2 \cite{supp}. Note that the expressions for the Lanczos coefficients are only valid in bulk and the edge corrections in $n \sim O(1)$ should be determined from the moments of the DOS \cite{Balasubramanian:2023kwd}. We numerically adjust the factor $\xi$ to make the analytical expression \eqref{bnpois} coincide with the value of $x$ where the numerical $\mathsf{b}(x)$ is maximum. We find excellent agreement with the numerical results with $\xi \simeq 1/2$, shown by the black dashed line (green region) in Fig.\,\ref{fig:lanczosinitialfull} (left).

Given the Lanczos coefficients in two extreme limits, we \emph{propose} the following \emph{Ansatz} to interpolate across the phases in the following form:
\begin{align}
    \mathsf{b}_{\gamma}(x)^2 = p\left(\frac{1 - x^{1 - q}}{1-q}\right) = - p \ln_q x\,, \label{anz1}
\end{align}
given by the $q$ logarithm, introduced by Tsallis \cite{Tsallis1994WhatAT}. It reduces to the usual logarithm for $q \rightarrow 1$, i.e., in the Poisson limit. Here, $p \equiv p(\gamma, N)$ and $q \equiv q(\gamma, N)$ weakly depend on the system size $N$. We scale the coefficients appropriately to lie between zero and unity. Importantly, \eqref{anz1} reduces to 
\eqref{bngoe} and \eqref{bnpois} at the $q \rightarrow 0$ and $q \rightarrow 1$ limits, which are proxy for the $\gamma \rightarrow 0$ and $\gamma \gtrsim 2$ limits, respectively. The corresponding limits of $p$ are $p \simeq 1$ and $p \simeq \xi^2/2$. Note that the \emph{Ansatz} \eqref{anz1} is not unique, and an alternate \emph{Ansatz} is discussed in the SM3 \cite{supp}.

Figure \ref{fig:lanczosinitialfull} (right) shows the dependence of $\gamma$ and $N$ on $p$ and $q$. The parameter $p \simeq 1$ in the ergodic regime, while close to the ergodic to fractal transition, starts to decrease, eventually saturating at $p \simeq \xi^2/2 \simeq 1/8$, which is an excellent agreement with the $\xi$ extracted from Fig.\,\ref{fig:lanczosinitialfull} (right). Conversely, the coefficient $q$ vanishes in the ergodic regime and eventually saturates at unity in the deep localized regime. The saturation values for $p$ and $q$ weakly depend on $N$ while converging to an asymptotic value in the large $N$ limit. This behavior signals a transition from ergodic to fractal properties close to $\gamma = 1$, where the parameters $p$ and $q$ are comparable. We compute the relative ``goodness of fit'' as $\varepsilon =\frac{\sqrt{\Delta p + \Delta q} \times 10^2}{\text{min}(p - 1, q -1) + 1}$ in SM2 \cite{supp}. This relative error is always $ < 10\%$, supporting the \emph{Ansatz} \eqref{anz1} as a good analytical form of the Lanczos coefficients across the full parameter regime of $\gamma$. Results for another quantity, the log-variance of Lanczos coefficients, are presented in SM4 \cite{supp}.

Curiously, the DOS corresponding to \eqref{anz1} can be obtained by solving the integral equation \eqref{den0}, which is valid in the bulk of the spectrum. We find
\begin{align}
    \rho_{\gamma}(E) = \sqrt{\frac{1}{4\pi p(1-q)}} \frac{\Gamma \big(\frac{1}{1-q}\big)}{\Gamma \big(\frac{1}{2}+\frac{1}{1-q}\big)} \Big(1-\frac{E^2}{4 p}(1 - q)\Big)^{\frac{1+q}{2(1- q)}}\,. \label{doe}
\end{align}
In the particular case of the $q \rightarrow 0$ and $q \rightarrow 1$ limits, \eqref{doe} reduces to the semicircle law and the Gaussian distribution \eqref{bnpois}, respectively, while \eqref{doe} holds across the full spectrum. It is normalized $\int_{-z}^z \rho_{\gamma}(E) \,dE = 1$ with $z = 2\sqrt{p/(1-q)}$. Both the limits suggest the effective variance as $p$, consistent with the observations in \cite{molina2005spectral, zuker2001canonical}. Moments of the above distribution can be straightforwardly computed; see SM5 \cite{supp}.

\emph{Behavior in the deep localized regime - analytic arguments:} A notable aspect of the RP model in its tridiagonal form is the persistence of nonzero off-diagonal elements even in the deeply localized regime, as evidenced by Fig.\,\ref{fig:lanczosinitialfull}. At first glance, this seems counterintuitive, especially when considering that at high values of $\gamma$, the primary contribution should theoretically arise from the diagonal matrix $A$. Yet, our numerical findings indicate that the off-diagonal elements remain finite and are of comparable magnitude to those found in the ergodic regime, a phenomenon we explore below.

We first examine the response of an orthogonal transformation to the RP model. For a Hamiltonian characterized by the distribution with $\langle H_{i j} \rangle = 0$ and $\langle H^2_{i j} \rangle = \alpha\delta_{i j} + \beta(1 - \delta_{i j})$, an orthogonal transformation via a matrix $C$ keeps the expectation value unchanged $\langle \tilde{H}_{i j} \rangle = 0$ while transforming the variance to $\langle \tilde{H}^2_{i j} \rangle = \beta(1 + \delta_{i j}) + (\alpha - 2 \beta)\sum_{l}c^2_{l i}c^{2}_{l j}$; see SM6 \cite{supp} for details. Here $\tilde{H}$ denotes the transformed Hamiltonian, and $c_{ij}$ represents the elements of $C$. Although this matrix can be arbitrary, we limit it to a special orthogonal matrix $M$ that executes the Householder transformation on $H$. The elements of $M$ can be approximated using the distribution of $H_{i j}$ \cite{supp}. The variance of the distribution of the elements of the matrix $\tilde{H}$ after a single orthogonal transformation is given by
\begin{align}
    &\langle \tilde{H}^2_{i j} \rangle \sim \beta (\delta_{i 1}+\delta_{j 1}) + \beta(1-\delta_{i j})\notag\\ &+ \frac{3}{N^2}(\alpha - 2\beta)(1-\delta_{i j})(1 - \delta_{i 1})
    + (\alpha - \beta)\delta_{i j}(1 - \delta_{i 1}) \notag\\ &+ \frac{1}{N}(\alpha - 2\beta)\left(\delta_{i 1}\delta_{j 1} + 2\delta_{i1} - 4 \delta_{i j} + 2\delta_{i j}\delta_{i 1}\right)\,.
\end{align}
These expressions result from a single step of the orthogonal transformation, assuming $\alpha, \beta \neq 0$. Cases where $\alpha = 0$ or $\beta = 0$ are considered separately. The complete tridiagonalization process involves $N$ such transformations, and these expressions are applied recursively to determine the final distribution of the Lanczos coefficients. The coefficients $\mathsf{b}(x)$ correspond to the norm of the off-diagonal terms $\tilde{H}_{1 j}$ at each step. We argue that the distribution of these off-diagonal terms is approximately Gaussian by the central limit theorem, or the ``law of large numbers''. This allows the norm to be estimated by the mean of the Nakagami distribution \cite{NAKAGAMI19603} with parameters $N$ and $\langle \tilde{H}^2_{1 j} \rangle$.

It is, thus, evident that although $\beta$ approaches zero deep in the localized regime, the full $\tilde{H}_{1 j}$ remains non-zero due to the $\alpha/N$ contribution (with $\alpha = 1$ for the RP model). This contribution has $1/N$ suppression, but this reduction is offset when calculating the norm of all $\tilde{H}_{1 j}$ (with $\langle \tilde{H}^2_{1 j} \rangle = \beta + \frac{2}{N}(\alpha - 2\beta)$) which introduces a $\sqrt{N}$ factor. Consequently, the overall suppression in the deeply localized regime is on the order of $N^{-1/2}$, persisting even at very high $\gamma$ values. This accounts for the collapse of Lanczos coefficients onto a single curve in the deeply localized regime, as shown in Fig.\,\ref{fig:lanczosinitialfull}.

\emph{Krylov IPR:} To gain further insight, we address the extension of localization in Krylov space by computing the inverse participation ratio (IPR) for the Krylov subspace. Instead of taking the usual eigenvectors, this amounts to considering the Krylov vectors corresponding to an initial localized state. It is widely understood that for quantum mechanical systems, Krylov space localization agrees with real space localization. The eigenvector IPR has been extensively studied in the RP model \cite{cadez2024rosenzweig}. We extend this study to the Krylov IPR and determine the appropriate scaling that allows us to distinguish between the three phases of the RP model. We define the Krylov IPR as the overlap between the computational basis state and the Krylov state
\begin{align}
    \mathrm{IPR}^{\ell}_\text{K}(\varphi_{k}) \equiv  \sum_{n = 1}^{N} \vert \langle n \vert \varphi_{k} \rangle \vert^{2 \ell} \,,
\end{align}
where $|\varphi_{k}\rangle$ stands for the $k^\text{th}$ Krylov vector and $\ket{n}$ is a computational basis element. The logarithm of the $\ell = 1$ case is often known as the \textit{participation entropy}. 

The Krylov state $\ket{\varphi_{k}}$ can be expressed in terms of the Hamiltonian eigenstates $\ket{\psi_m}$ as $\ket{\varphi_{k}} = \sum_{m = 1}^{N}\eta^{k}_{m}\ket{\psi_m}$. The coefficients $\eta^{k}_{m}$ satisfy a three-term recursion relation; see SM7 \cite{supp}. Thus, $\mathrm{IPR}^{\ell}_\text{K}$ can be further rewritten as $ \mathrm{IPR}^{\ell}_\text{K}(\varphi_{k}) =   \sum_{n = 1}^{N} \Big\vert \sum_{m = 1}^{N}\eta^{k}_{m}s^{m}_{n} \Big\vert^{2 \ell}$ where $s^{m}_{n} = \braket{ n \vert \psi_{m}}$. The usual IPR for the eigenstate $|\psi_{k}\rangle$ is then obtained by simply setting $\eta_m^k = \delta^{m}_{k}$ i.e., the usual eigenstate IPR information is encoded in the Krylov IPR. In SM7 \cite{supp}, the overlap of Krylov vectors and eigenstates is also investigated. This indicates that the Krylov IPR can serve as a robust indicator for both localization and multifractality \cite{khaymovich2021dynamical}. Physically, this corresponds to the spread of eigenstates on the computational basis and the spread of the Krylov vectors in the eigenbasis. Note that our definition of the Krylov IPR is \emph{static} and reflects the localization with respect to the $\gamma$ in comparison to the \emph{dynamic} IPR defined in \cite{Bento:2023bjn}.

\begin{figure}[t]
\hspace*{-0.5cm}
\includegraphics[width=0.5\textwidth]{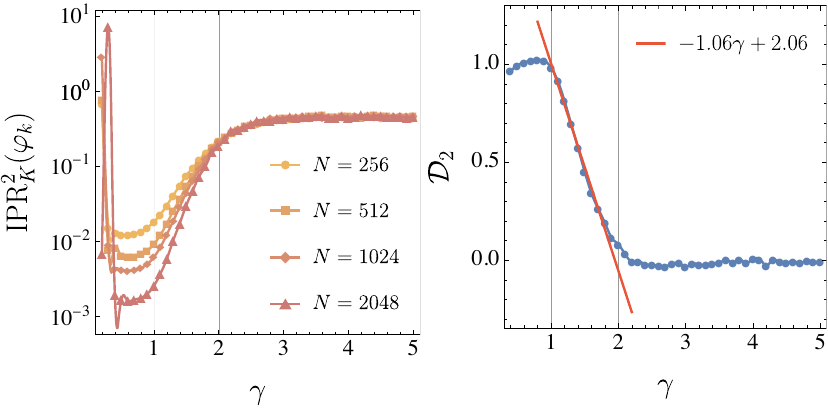}
\caption{Left: Scaling of the $\mathrm{IPR}^{2}_{\text{K}}$ for $\varphi_{N-1}$ for different values of $N$ as a function of $\gamma$. The scaling with system size is nearly unaffected by $\gamma$. The numerical instability at low $\gamma$ arises due to the finite size of the system, which reduces with increasing system size. Right: Scaling exponent $\mathcal{D}_{2}$ of $\mathrm{IPR}^{2}_{\text{K}}$ for $\varphi_{N - 1}$ as a function of $\gamma$. In both cases, the exponent is nearly constant in the ergodic regime and decreases linearly in the fractal regime to settle at zero in the localized regime.} \label{fig:kiprcomp}
\end{figure}

The detailed behavior of the Krylov IPR is discussed in the SM7 \cite{supp}. Here, we are interested in the scaling of the IPR, which is much more apparent for $k = N - 1$ than $k  = N/2$, i.e., the middle of the Krylov spectrum. The results are presented in Fig.\,\ref{fig:kiprcomp}, and we compare them with the $\ell=1$ case in SM7 \cite{supp}. The IPR is higher in the localized regime compared to the ergodic and fractal phases. The scaling exponent is obtained by fitting the linear decrease $\mathrm{IRP}^{2}_{\text{K}}(\varphi_{N - 1}) \sim N^{-\mathcal{D}_2}$ in the fractal region. We observe the following features of the scaling exponent
\begin{align}
    \mathcal{D}_{2}(\gamma) \sim \begin{cases}
     1 & ~~\gamma \leq 1\,,\\
    2 - \gamma &~~ 1 < \gamma \leq 2\,,\\
    0 & ~~\gamma > 2\,.
  \end{cases}\label{D2}
\end{align}
Up to the numerical and the finite-size effects, $\mathcal{D}_{2}(\gamma) = \text{constant} > 0$ for $\gamma \leq 1$ (ergodic phase). In the fractal phase, it linearly decreases $\mathcal{D}_{2}(\gamma) \sim -\alpha \gamma + \chi > 0$ (where $\alpha$ and $\chi$ are some positive constants) for $1 < \gamma \leq 2$ (fractal phase) while $\mathcal{D}_{2}(\gamma) \sim 0$ for $\gamma > 2$ (localized phase), indicating strong localization. 

The underlying reason behind the strong sensitivity of $\varphi_{N-1}$ (or any $\varphi_{k}$ close to $k = N$) to $\gamma$ comes from the Lanczos algorithm. In the context of matrix diagonalization, the Lanczos algorithm converges from both extreme ends, making it particularly effective for finding the smallest or largest few eigenvalues and eigenvectors. However, to obtain the eigenvalues and eigenvectors in the middle of the spectrum, the algorithm must effectively span the entire basis. Since $\varphi_{k}$ is generated by the application of $H^{k}$ on the initial state followed by the subtraction of the contribution of $\varphi_{k - 1}$ and $\varphi_{k - 2}$, the effect of the full Lanczos spectrum is encoded in the Krylov vectors toward the end of the spectrum. For example, by considering the overlap of all Krylov vectors with the mid-spectrum eigenstate, one finds that the larger the value of $k$, the more information about the Lanczos spectrum is encoded in this overlap coefficient. Since we have seen that the full Lanczos spectrum is sensitive to the transition, it is desirable to utilize a Krylov vector that encodes the full spectrum. This also explains why $\varphi_{N-1}$ is highly sensitive to mid-spectrum eigenstates.

\begin{figure}[t]
\hspace*{-0.5cm}
\includegraphics[width=0.45\textwidth]{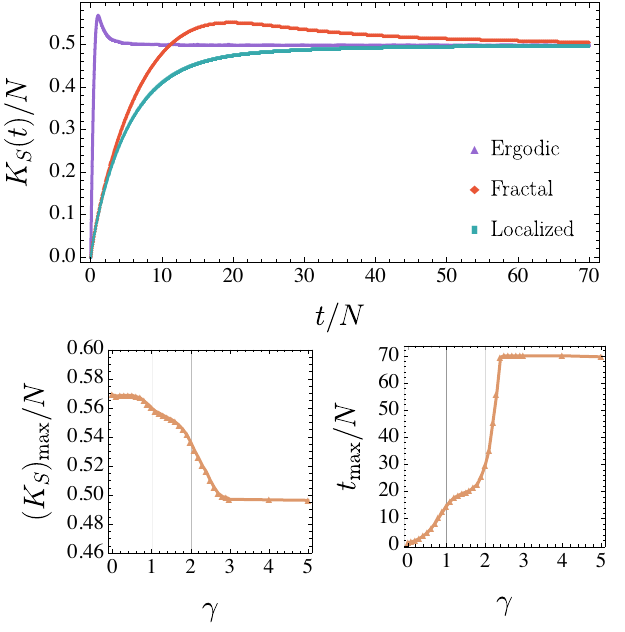}
\caption{Top: Spread complexity for the RP model \eqref{RPmodel} in different regimes, initialized by a TFD state at infinite temperature. Bottom, left: The variation of the peak value of the spread complexity, and (bottom right) the peak time in different phases of the Hamiltonian \eqref{RPmodel}, initialized by a TFD state at infinite temperature. The system size is $N = 500$, variance $\sigma^2 = 1/N$ with $800$ Hamiltonian realizations.} \label{fig:KtspreadTFD}
\end{figure}

\emph{Spread complexity:} The rate of spreading of the information in different phases can be captured by the spread complexity associated with the Hamiltonian \eqref{RPmodel}, initializing with the maximally entangled infinite-temperature TFD state. As depicted in Fig.\,\ref{fig:KtspreadTFD}, the spread complexity exhibits distinct behaviors across three regimes: ergodic (purple), fractal (red), and localized (teal). Notably, pronounced peaks are observable in the ergodic and fractal regimes, contrasting their absence in the localized regime. This disparity suggests that the emergence of peaks, and consequently a linear slope, may be attributed to the level-repulsion among the eigenvalues \cite{Balasubramanian:2022dnj, Erdmenger:2023wjg, Camargo:2023eev, Camargo:2024deu}.

The saturation value of spread complexity always converges to a value of $\Bar{K}_S/N = (N-1)/(2N) \approx 0.5$ \cite{Erdmenger:2023wjg}. In contrast, the peak value of spread complexity exhibits a gradual decline as we transition into the localized regime. In the latter case, the peak is notably absent, rendering the peak value synonymous with the saturation value. Analogously, the peak time of spread complexity demonstrates a clear differentiation in growth across the three regimes. This is marked by transitions at the critical points of $\gamma = 1$ and $\gamma = 2$, highlighting the unique dynamics within each regime.

\emph{Conclusion and outlook:} In this study, we investigate the Rosenzweig-Porter (RP) model, a random matrix model that exhibits extended, fractal, and localized phases. Our exploration utilizes static and dynamic measures within the Krylov subspace to illuminate the phase transitions among these three distinct states. We present a semianalytic expression for the Lanczos coefficients that interpolates between the three phases. We introduce the Krylov inverse participation ratio (IPR) and a Krylov fractal dimension, which accurately indicate the two critical transition points. Notably, the \textit{edge-spectrum} Krylov vectors exhibit stronger sensitivity to fractality, in contrast to those near the center of the spectrum. Dynamic metrics, such as the spread complexity of Krylov states, also reflect these transitions.

Our findings affirm that both static and dynamic Krylov subspace probes are attuned to the transitions from ergodicity through fractality to localization. These probes present robust alternatives to eigenspectrum analysis. This work bridges a significant gap in the literature on Krylov techniques, especially by providing an analytic expression of Lanczos coefficients across the full parameter regime, thereby establishing a standard for detecting transitions from chaotic to nonchaotic states in physical systems via Krylov methodologies \cite{Nandy:2024htc}.

An immediate extension of this method could be the utilization of these probes in systems undergoing localization transitions. The behavior of the DOS \eqref{doe} shares similar features in two extreme limits of the spin-glass \cite{erdHos2014phase} or the double-scaled SYK model \cite{Berkooz:2018qkz} (see SM for some initial attempts). It would be intriguing to assess how the proposed \emph{Ansatz} \eqref{anz1} compares to the tridiagonal elements found in the models with corresponding $q$-normal distribution \cite{erdHos2014phase, Berkooz:2018qkz}. This seems to be illuminating if one computes the moments of the Hamiltonian \eqref{RPmodel} using the chord diagrams \cite{Zhuothanks}, where the $A$ and $B$ matrices provide the intersecting and the nonintersecting chords. Future research might explore such transitions within various matrix ensembles and different diagonal distributions \cite{replicaScipost}. We anticipate that the overarching characteristics will align with our findings. These probes could also be applied to non-Hermitian extensions within the 38-fold symmetry classes \cite{PhysRevX.9.041015}, particularly those lacking a fractal phase in generic complex diagonal Hamiltonians \cite{detomasi2022nonhermitian}. In generic systems, extracting the exact form of the Lanczos coefficients across the full spectrum with a known DOS is also unknown. We sketch a general methodology in SM5 that may benefit such endeavors.

\emph{Note added:} Recently, we became aware of an upcoming work \cite{Dymupcoming} that also obtains the same logarithmic function \eqref{bnpois} in the Gaussian limit.

\emph{Acknowledgments:} We thank Tanay Pathak for the initial collaboration and feedback. We also acknowledge valuable discussions with Alexei Andreanov, Adolfo del Campo, Barbara Dietz, Anatoly Dymarsky, Andr\'as Grabarits, Ivan Khaymovich, Rafael Molina, Horacio Pastawski, Anatoli Polkovnikov, Dario Rosa, Lea Santos, Ruth Shir, Luca Tessieri, and Zhuo-Yu Xian. B.B. thanks the organizers of the International Workshop of Disorder Systems (Salamanca) 2024 and the University of Luxembourg for their hospitality during the final stages of the preparation of this Letter. P.N. thanks the University of Kentucky and the Berkeley Center for Theoretical Physics, University of California, Berkeley, for hosting him through the Adopting Sustainable Partnerships for Innovative Research Ecosystem (ASPIRE) program of the Japan Science and Technology Agency (JST), Grant No. JPMJAP2318, during the final stages of the work. B.B. acknowledges financial support from the Institute for Basic Science (IBS) in the Republic of Korea through Project No. IBS-R024-D1. The work of P.N. is supported by the Japan Society for the Promotion of Science (JSPS) Grant-in-Aid for Transformative Research Areas (A) “Extreme Universe” No. JP21H05190. 
\bibliography{references}

\newpage

\hbox{}\thispagestyle{empty}\newpage
\appendix\label{appendix}
\onecolumngrid
\begin{center}
\textbf{\large Supplemental Materials (SM): Krylov fractality and complexity in generic random matrix ensembles}
\end{center}

\section{SM1: Level statistics of the RP model}

\begin{figure}[h]
\hspace*{-0.7cm}
\includegraphics[width=0.42\textwidth]{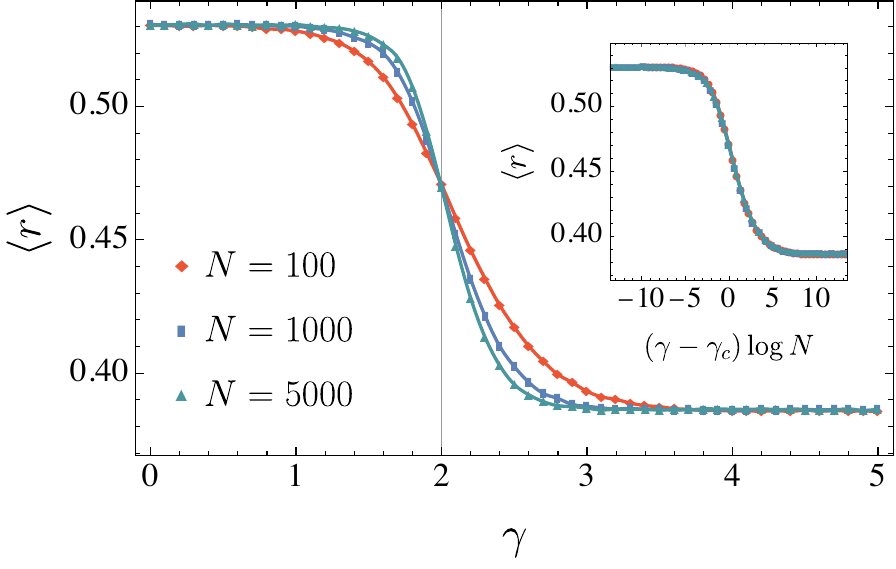}
\caption{(Main) The $\braket{r}$ value statistics with varying $\gamma$ for the RP model for different system sizes, $N = 100 \,(50000), N = 1000\, (5000)$, and $N = 5000 \,(500)$, where the respective number of averages are mentioned in the brackets. A clear transition is visible at $\gamma_c = 2$. (Inset) The same  $\braket{r}$ value statistics against $(\gamma - \gamma_c) \log N$ to exhibit the data collapse.} \label{fig:rprval}
\end{figure}

The Bohigas-Giannoni-Schmidt (BGS) conjecture tells us that quantum chaotic systems are characterized by level repulsion, which leads to Wigner-Dyson statistics \cite{Wigner1, Dyson1962a} in their eigenspectrum \cite{BGS}. On the other hand, the Berry-Tabor conjecture suggests that integrable are characterized by level crossings, leading to Poisson statistics \cite{Berry_Tabor_conj}. In numerical computations, it is often advantageous to compute the $\braket{r}$, which circumvents the unfolding procedure necessary to obtain the level statistics. The $\braket{r}$ value is defined as the mean value of the consecutive level spacing ratio 
\begin{align}
    r_n = \frac{\mathrm{min}(s_{i}, s_{i+1})}{\mathrm{max}(s_{i}, s_{i+1})}\,,~~~ \braket{r} = \mathrm{mean} (r_n)\,.
\end{align}
Here $s_n = E_{n+1} - E_n$ denotes the consecutive level spacing. Figure \ref{fig:rprval} shows the  $\braket{r}$ value with respect to $\gamma$ for different system sizes of the RP model \eqref{RPmodel}. A sufficient number of realizations of individual Hamiltonian is taken. Clearly, it does not show any transition at $\gamma = 1$, hence the short-range level statistics is blind to the transition between the ergodic and the fractal regime. In contrast, it shows a crossover at $\gamma = 2$ which is the boundary between the fractal and the localized regime. The inset shows the data collapse which makes the crossover more apparent at $\gamma = 2$.

\section{SM2: Structure of the inverse binomial function}

We briefly sketch how to obtain \eqref{bnpois} from \eqref{nib1}. To obtain the form of the ``nib'' function around $x \sim 0$, we let
\begin{align}
    g = \mathrm{nib}(x,d)\,.
\end{align}
This implies the inverse is the binomial function $x = \mathrm{bin}(g,d)$. Using, the definition of the shifted binomial \cite{zuker2001canonical}, one can write \cite{zuker2001canonical}
\begin{align}
    x =  \mathrm{bin}(g,d) = 2^{-d} \binom{d}{d(1/2-g)} \approx e^{-\frac{d}{2}[(1+2g)\log (1+2g) + (1-2g) \log(1-2g)]} \simeq e^{-2 d g^2}\,,
\end{align}
where third equality is obtained by the Stirling approximation to the Binomial coefficients \cite{MacKaybook}
\begin{align}
    \ln \binom{n}{r} \simeq (n-r) \ln \frac{n}{n-r} + r \ln \frac{n}{r}\,,
\end{align}
and last equality comes from the dominant contribution of the saddle point located at $g = 0$ in the large $d$ limit. Hence, inverting, we obtain
\begin{align}
    g = \mathrm{nib}(x,d) \simeq \sqrt{-\frac{1}{2 d} \ln x}\,,
\end{align}
from which \eqref{bnpois} follows.

\section{SM3: Alternate ansatz for Lanczos coefficients}

\begin{figure}[t]
\centering
\includegraphics[width=0.42\textwidth]{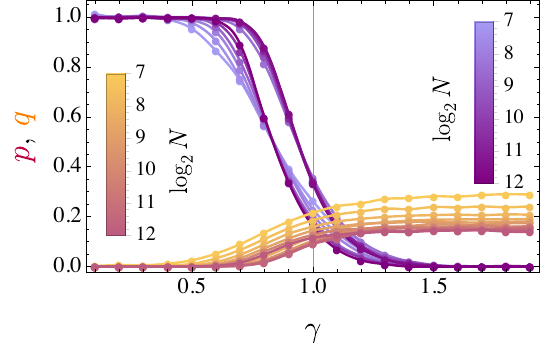}
\label{fig:lanczosinitialfit2}
\caption{The variation of $p$ and $q$ for the \emph{Anstaz} \eqref{anz2}. Compare this to Fig.\,\ref{fig:lanczosinitialfull} (right) for the \emph{Anstaz} \eqref{anz1}.} \label{fig:lanczosinitialfull2}
\end{figure}

In the main text, we considered an \emph{Ansatz} \eqref{anz1} for the Lanczos coefficients across the full spectrum. Here we present another \emph{Ansatz} which can also be chosen as an alternative to \eqref{anz1}. We consider the following \emph{Ansatz}:
\begin{align}
    \mathsf{b} (x)^2 = p(\gamma, N)(1-x) - q(\gamma, N)\ln x\,, \label{anz2}
\end{align}
as a superposition of the two limiting cases in \eqref{bngoe} and \eqref{bnpois}. Since the logarithmic function is ill-defined, this \emph{Ansatz} does not strictly hold at $x = 0$. Barring this condition, the bulk is well approximated by a numerical fit. Additionally, the ergodic to fractal transition point is characterized by the following form
\begin{align}
    \mathsf{b}(x)^2\vert_{\gamma = 1} = w(N) ( 1 - x - \ln x)\,,
\end{align}
where the proportionality constant $p = q = w(N)$ weakly depends on $N$. We fit \eqref{anz2} to the numerical Lanczos coefficients and find the functions $p$ and $q$. Figure \ref{fig:lanczosinitialfull2} shows the behavior of $p$ and $q$. The function $p$ starts at unity in the ergodic regime, followed by a decrease near $\gamma = 1$, and finally vanishes for $\gamma \lesssim 2$. The function $q$, on the other hand, vanishes in the fully ergodic phase and starts to increase near $\gamma = 1$ saturating at $\gamma \sim 2$. The ergodic to fractal transition is the approximate region where $p$ and $q$ become comparable.

\begin{figure}[t]
\centering
\includegraphics[width=1\textwidth]{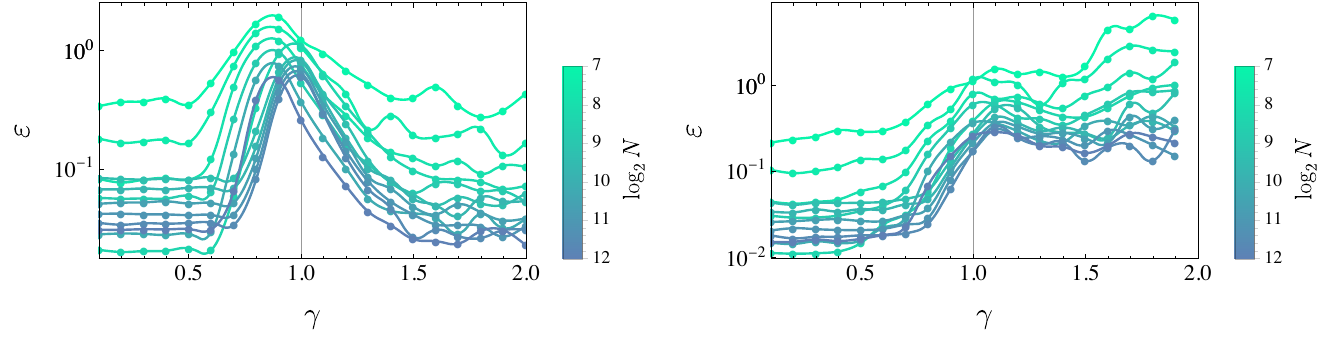}
\caption{((a) The goodness of fit (as $\%$ error) as a function of $\gamma$ and system size $N$ in the original ansatz. (b) The goodness of fit (as $\%$ error) as a function of $\gamma$ and system size $N$ in the alternative ansatz.} \label{fig:fitab2}
\end{figure}

Figure \ref{fig:fitab2} shows the relative ``goodness of fit'' introduced in the main text for both the \emph{Ans\"atze}. The error peaks the maximum around $\gamma = 1$, the transition between the ergodic to the fractal regime. The numerical results, presented in Fig.\,\ref{fig:lanczosinitialfull2} and Fig.\,\ref{fig:fitab2} corroborate the same and indicate that the ansatz \eqref{anz2} is reasonably accurate.

\section{SM4: Log-variance of the Lanczos spectrum}

\begin{figure}[t]
\centering
{\includegraphics[width=0.43\textwidth]{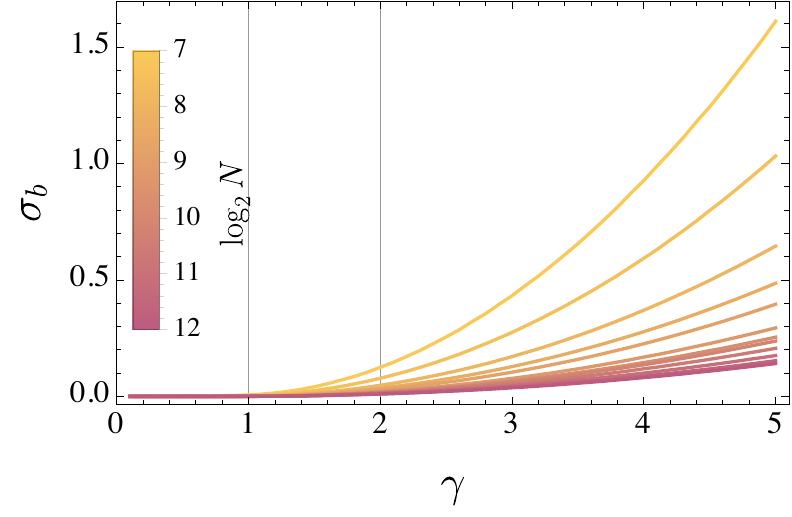}}
\hfill
{\includegraphics[width=0.47\textwidth]{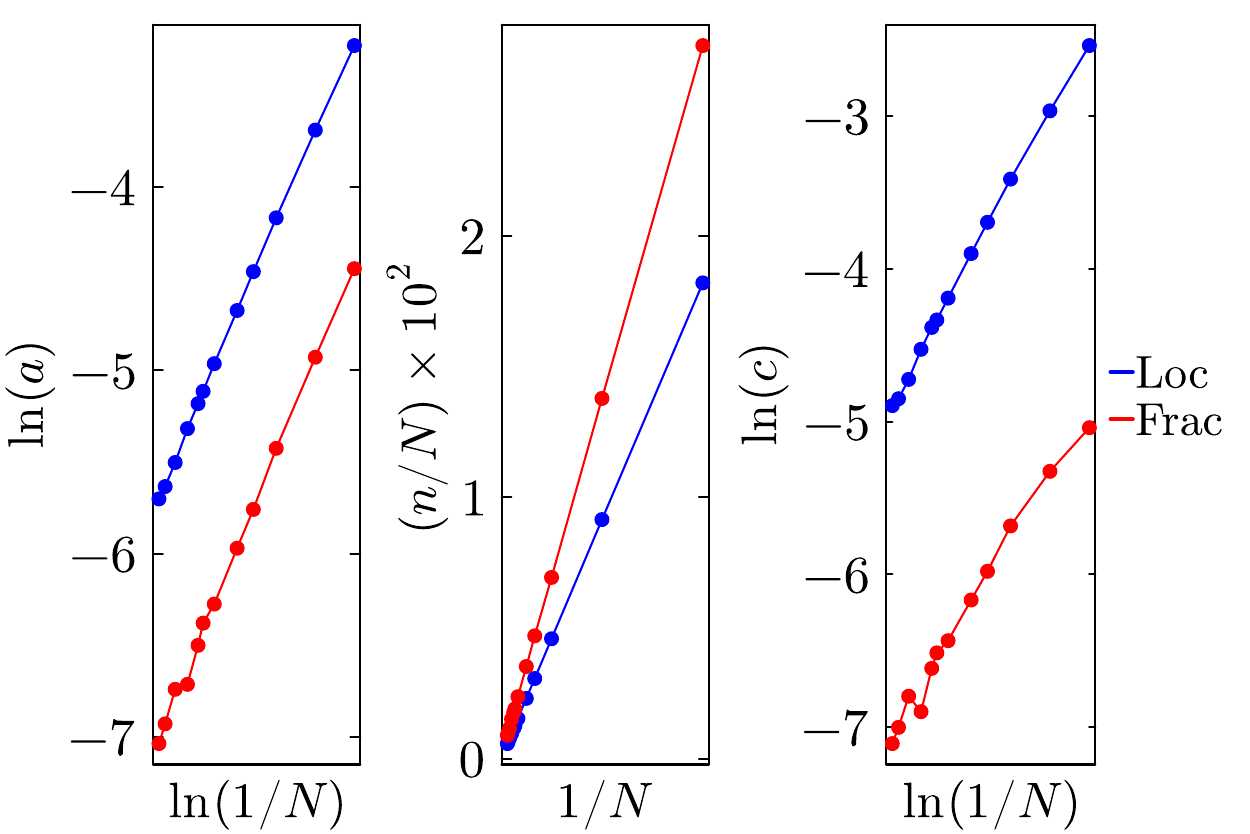}}
\caption{(Left) Log-variance of the Lanczos coefficients of the RP model \eqref{RPmodel} as a function of parameter $\gamma$ for different system sizes $N$. (Right) The variation of the parameters $a(N)$, $n(N)$, and $c(N)$ with the appropriate function of $N$, in the fractal and localized phases.} \label{fig:rplogvar}
\end{figure}

We further investigate the Krylov spectrum to quantify the two transitions in Krylov space. The probe we consider numerically is the log-variance of the Lanczos spectrum \cite{Rabinovici:2021qqt}. This gives us information about the spread of the spectrum around its mean value and is in general understood to capture the ergodic-localized transition. We study its behavior in the multifractal regime and see how that changes across the two transition points. We consider $12$ system sizes and study how the log-variance, defined as \cite{Rabinovici:2021qqt}
\begin{align}
    \sigma_{b} = \text{Var}\left(\ln \frac{b_{2 j - 1}}{b_{2 j}} \right)\,,
\end{align}
as a function of the parameter $\gamma$. The results presented correspond to averaging over many realizations. It is expected that the variance will be low in the ergodic regime and higher in the non-ergodic regimes \cite{Rabinovici:2021qqt}. This is observed for the RP model and is presented in Fig.\,\ref{fig:rplogvar} (left). It is evident that \(\sigma_b\) nearly vanishes in the ergodic regime \(\gamma < 1\) and begins to grow upon ergodicity breaking for \(\gamma > 1\). Naturally, the growth is slower for larger system sizes since the effective ergodicity breaking parameter decreases with increasing system size.

It is instructive to determine the exact nature of \(\sigma_{b}\) as a function of $\gamma$ and $N$. To do this, we fit our numerical results to the following power-law \emph{Ansatz}
\begin{align}
    \tilde{\sigma}_{b}(\gamma, N) = a(N)\gamma^{n(N)} + c(N)\,.
\end{align}
The respective parameters $a, n$, and $c$ are presented in Fig.\,\ref{fig:rplogvar} (right).
From the data, it is straightforward to read off the dependence of the respective parameters on $N$ in the two phases as follows
\begin{align}
    \ln (a)_{\text{Frac}} &= -0.7485\ln (N) - 0.7853 \,,~~~~~
    n_{\text{Frac}} = 3.4910 + 8.10\times 10^{-5} \times N\,,\\
    \ln (a)_{\text{Loc}} &= -0.7208\ln (N)  + 0.3081\,,~~~~~
    n_{\text{Loc}} = 2.3312 + 1.90\times 10^{-5} \times N  \,.
\end{align}
The functional form of $c(N)$ is harder to determine. However, it is not as relevant to the behavior of $\tilde{\sigma}_{b}$ as $a(N)$ and $n(N)$. From the fitting data, it is evident that the two phases (fractal and localized) demonstrate different behavior. Therefore, we conclude that the behavior of the log-variance probes the ergodic, fractal, and localized phases. However, the system-size dependence is quite strong and the difference between the fractal and localized phases may be lost in smaller system sizes. Thus, the log-variance is a much stronger probe of the ergodic to fractal transition (since it transforms from zero to a non-zero value) as compared to the fractal to localized transition.

\section{SM5: Moments of the distribution \eqref{doe}}

\begin{figure}[t]
\centering
\includegraphics[width=1\textwidth]{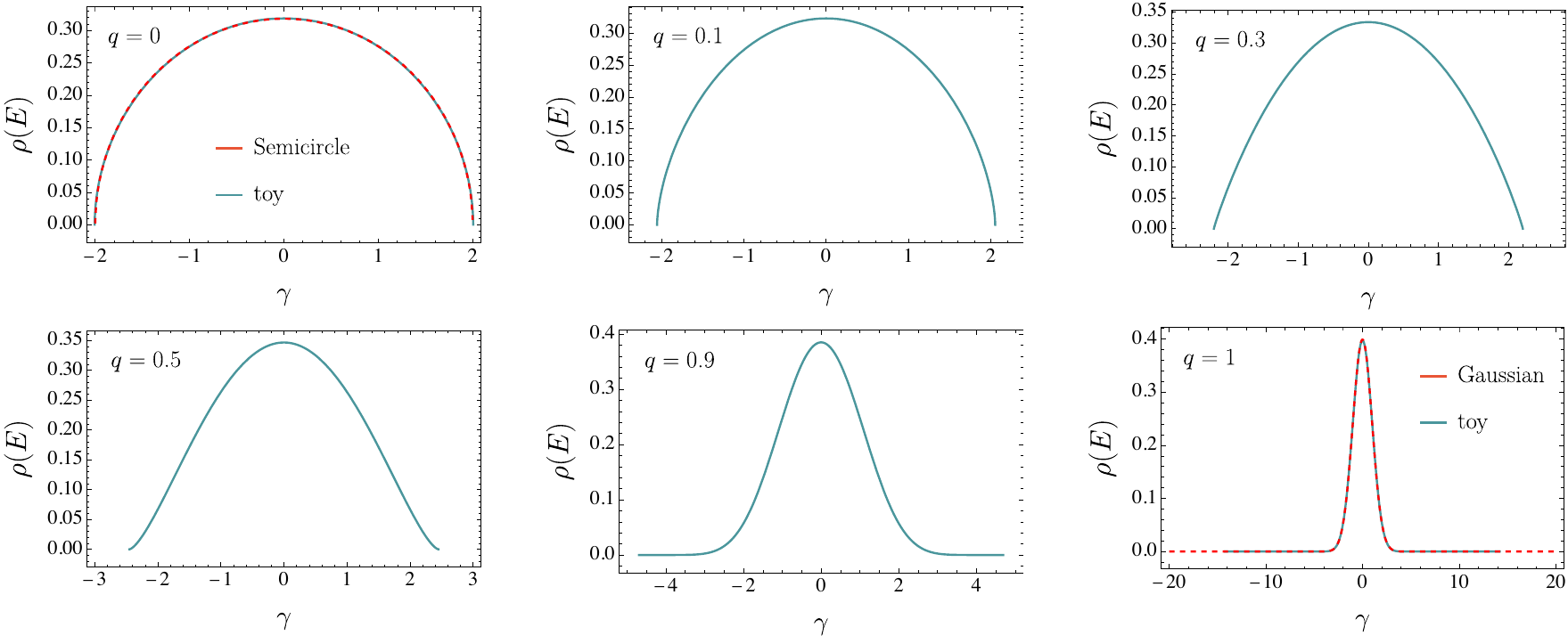}
\caption{The density of the states (DOS) \eqref{doea} for different $q$ is shown. For both cases, the extreme limits $q = 0$ and $q =1$, match perfectly for the semicircle and the Gaussian distribution.} \label{fig:toy}
\end{figure}

We briefly discuss the properties of the distribution of the moments of the DOS \eqref{doe}. For convenience, we choose $p = (1- q/2)$, which interpolates between the semicircle and the normal distribution. However, a more general form $p = (1- f(q)/2)$ can be chosen such that $f(0) = 0$ and $f(1) = 1$. With the above choice, the Lanczos coefficients are given by
\begin{align}
    \mathsf{b}(x)^2 = \left(1-\frac{q}{2}\right)\left(\frac{1 - x^{1 - q}}{1-q}\right) = - \left(1-\frac{q}{2}\right) \ln_q x\,, \label{anz1s}
\end{align}
with $q \equiv q (\gamma)$. For $q = 0$ and $q=1$, it reduces to the corresponding semicircle and the Gaussian limits. For the generic form of \eqref{anz1s}, the DOS is given by
\begin{align}
    \rho_{\mathrm{toy}}(E) = \sqrt{\frac{1}{2\pi (2-q)(1-q)}} \frac{\Gamma \big(\frac{1}{1-q}\big)}{\Gamma \big(\frac{1}{2}+\frac{1}{1-q}\big)} \Big(1-\frac{(1 - q)}{2 (2-q)}E^2\Big)^{\frac{1+q}{2(1- q)}}\,. \label{doea}
\end{align}
The DOS is normalized such that $\int_{-u(q)}^{u(q)} \rho(E)\, dE = 1$ with $z(q) = \sqrt{2(2-q)/(1-q)}$. See Fig.\,\ref{fig:toy} for the behavior of DOS for different $q$. We are interested in the moments of the DOS\footnote{These are proportional to the moments of the Hamiltonian in the given initial state, provided the state has a smooth and nearly uniform dependence on the eigenstates \cite{Balasubramanian:2023kwd}.}:
\begin{align}
    m^{\mathrm{toy}}_n(q) = \int_{-z(q)}^{z(q)} E^n \, \rho_{\mathrm{toy}}(E)\, dE\,.
\end{align}
Due to the symmetry of $\rho_{\mathrm{RP}}(-E) = \rho_{\mathrm{RP}}(E)$, the odd moments vanish. The even moments can be straightforwardly calculated and are given by
\begin{align}
    m^{\mathrm{toy}}_{2n}(q) = (2 n-1)\text{!!} \,\frac{(2-q)^n }{(1-q)^{n+1} } \frac{\Gamma \big(\frac{1}{1-q}\big)}{\Gamma \big(n + 1+\frac{1}{1-q}\big)}\,. \label{momrp}
\end{align}
Since $0\leq q < 1$, the denominator never vanishes. The moments in the two extreme limits are given by
\begin{align}
    \lim_{q \rightarrow 0} m^{\mathrm{toy}}_{2n}(q) = \frac{2^n (2 n-1)\text{!!}}{\Gamma (n+2)} = \frac{(2n)!}{n! (n+1)!}\,,~~~~~~~ \lim_{q \rightarrow 1} m^{\mathrm{toy}}_{2n}(q) = (2 n-1)\text{!!} = \frac{(2n)!}{2^n n!}\,.
\end{align}
From the above moments \eqref{momrp}, we employ the iterative algorithm \cite{viswanath1994recursion, Parker:2018yvk, Nandy:2024htc} to compute the lower-order Lanczos coefficients:
\begin{align}
    a^{\mathrm{toy}}_n(q) = 0\,, ~~~~~~~~  b^{\mathrm{toy}}_n (q) = \sqrt{\frac{(2-q) n \left(\frac{n+1}{2}-\frac{n-1}{2}  q\right)}{[(n+1)-n q]\, [n-(n-1) q]}}\,
\end{align}
Since $0\leq q < 1$, the denominator is always non-vanishing. In the two extreme limits, we have
\begin{align}
    \lim_{q \rightarrow 0} b^{\mathrm{toy}}_{n}(q) = 1\,, ~~~~~~ \lim_{q \rightarrow 1} b^{\mathrm{toy}}_{n}(q) = \sqrt{n}\,.
\end{align}
For finite-size systems, such lower-order coefficients can be matched with the numerical Lanczos coefficients with an initial state that has smooth dependence on the energy basis \cite{Balasubramanian:2023kwd}. We leave such study for future work.



\section{SM6: Tridiagonalizing Rosenzweig-Porter model}
\label{app2}

Here we present the details of the analytical arguments for the approximation of the tridiagonal form of the RP model, briefly discussed in the analytical arguments. The following subsections present the steps sequentially and at the end, we present an algorithm that can be used to determine the approximate distribution of the Lanczos coefficients. 
\subsubsection{Orthogonal transformation}
We begin by considering the following form of a Gaussian matrix
\begin{align}
    \langle H_{i j} \rangle = 0\,,~~~~~
    \langle H^2_{i j} \rangle = \alpha\delta_{i j} + \beta(1 - \delta_{i j})\,. \label{hetm}
\end{align}
A special case of this general Gaussian matrix corresponds to the RP model with $\alpha = 1/(2N)$ and $\beta = 1/(4 N^{\gamma + 1})$. This type of matrix is known as a Heteroskedastic matrix, where the defining feature is different variances for different elements.  The specific Heteroskedastic matrix we consider is one where all the off-diagonal elements have the same variance, while all the diagonal elements have a different variance. For such a matrix, the action of an orthogonal transformation $C$ gives the following result
\begin{align}
    \tilde{H}_{i j} = (C^{T}H C)_{i j} = \sum_{l, k}c_{l i}c_{k j}H_{l k}\,.\label{hij}
\end{align}
It is straightforward to see that $\langle \tilde{H}_{i j} \rangle = 0$. The variance $\langle \tilde{H}^2_{i j} \rangle$ can be similarly computed to give the following result
\begin{align}
    \langle \tilde{H}^2_{i j} \rangle = \beta(1 + \delta_{i j}) + (\alpha - 2 \beta)\sum_{l}c^2_{l i}c^{2}_{l j}\,. \label{ortho1}
\end{align}
Note that the variances now depend on the exact details of the orthogonal matrix $C$ \textit{unless} $\alpha = 2 \beta$, which is the case for random Gaussian Wigner matrices. Since in our general case this is not true, we are forced to make some approximations. We shall approximate the magnitude of $\sum_{l}c^2_{l i}c^{2}_{l j}$ by resorting to a particular construction of $C$. In order to motivate the form of $C$, we review the tridiagonalization procedure for usual random symmetric GOE matrices. 
\subsubsection{Tridiagonalizing GOE matrices}

To tridiagonalize a random symmetric GOE matrix $H_{N}$ of dimension $N$, we first write the matrix as follows \cite{dumitriu2002matrix}
\begin{align}
 H_{N} =   \begin{pmatrix}
  a_0 & v^{T}\\ 
  v & H_{N - 1}
\end{pmatrix}\,,
\end{align}
with $H_{i i } \sim \mathcal{N}(0,2)$ and $H_{i j} \sim \mathcal{N}(0,1)$. Here $a_0$ is a diagonal element of the GOE matrix, taken from the distribution $\mathcal{N}(0,2)$. The $N - 1$ dimensional vector $v$ have components taken from $\mathcal{N}(0,1)$. We start with the following orthogonal matrix
\begin{align}
    O = \begin{pmatrix}
  1 & 0^T_{N - 1}\\ 
  0_{N - 1} & M_{N - 1}
\end{pmatrix}\label{orthoO}\,,
\end{align}
where $0_{N - 1}$ represents an $N - 1$ dimensional vector with all elements $0$. The $N - 1$ dimensional matrix $M_{N - 1}$ is defined in the following way
\begin{align}
    M . v = \vert \vert v \vert \vert e_{1}\,,
\end{align}
where $e_{1} = (1,0,\dots,0,0)_{N - 1}$ and $\vert \vert v \vert \vert$ is the norm of the vector $v$. This ensures that the matrix $H$ takes the following form under the orthogonal transformation $O^T H O$
\begin{align}
    O^T H O =  \begin{pmatrix}
  a_0 & \vert \vert v \vert \vert e^T_{1}\\ 
  \vert \vert v \vert \vert e_{1} & M^T H_{N - 1} M
\end{pmatrix}\,.
\end{align}
This matrix is tridiagonal in the first row and column. Since $H_{N - 1}$ is a symmetric GOE matrix, it remains a symmetric GOE matrix after the orthogonal transformation $M^T H_{N - 1} M$. Therefore the same process can be repeated by choosing the following orthogonal matrix
\begin{align}
    Q = \begin{pmatrix}
  1 & 0^T_{(N - 1)/2} & 0^T_{(N - 1)/2}\\ 
  0_{(N - 1)/2} & 1 & 0^T_{N - 2} \\
  0_{(N - 1)/2} & 0_{N - 2} & S_{N - 2} 
\end{pmatrix}\label{matQ}\,,
\end{align}
and evaluating $Q^T O^T H O Q$. This matrix is tridiagonal in the first two rows and columns. 

This process is repeated $N - 1$ times and a full tridiagonal matrix is obtained. The distribution of the diagonal elements remains the same as the original matrix, i.e. $\mathcal{N}(0,2)$. The off-diagonal elements correspond to the norms of the vectors that make up the rows and columns of the upper/lower-triangular parts of the matrix $H_{N}$. These are given by the Chi-distribution $\{\chi_{N - 1}, \chi_{N - 2}, \dots, \chi_{1}\}$.

\subsubsection{Householder matrix}
Our main takeaway from this calculation is the form of the matrix $M$. This is known as the Householder transformation, and while it takes many forms, we use the following form \cite{dumitriu2002matrix}
\begin{align}
    M = \mathbb{I} - 2\frac{u^{T}u}{u u^{T}}\,,~~~~ u = v - \vert \vert v \vert \vert e_{1}\,.
\end{align}
To utilize this, we require the probability distribution of $\vert \vert v \vert \vert$. The norm of an $L$ dimensional Gaussian random vector with mean $0$ and variance $\sigma$ is given by
\begin{align}
    P_{L, \sigma}(x) = \frac{2^{1-\frac{L}{2}}x^{L-1} e^{-\frac{x^2}{2 \sigma ^2}}}{\Gamma \left(\frac{L}{2}\right) \sigma^{L}}\,.
\end{align}
This is the Nakagami distribution $f(m, \Omega)$ with the parameters $(m, \Omega) = (L/2, L \sigma^2)$ \cite{NAKAGAMI19603}. For simplicity, we shall use the following shorthand to denote the Nakagami distribution $f(L/2, L\sigma^2)) \equiv V_{L}(\sigma)$. We shall use this distribution, along with the Gaussian distribution, to estimate the size of the elements of $M$. 

The elements of the matrix $M$ are given by the following relation (for a fixed vector $v$)
\begin{align}
    M_{i j} = \delta_{i j} - \frac{v_i v_j}{\vert \vert v \vert \vert^2 - \vert \vert v \vert \vert v_1} + \frac{\vert \vert v \vert \vert v_i \delta_{j 1}}{\vert \vert v \vert \vert^2 - \vert \vert v \vert \vert v_1} + \frac{\vert \vert v \vert \vert v_j \delta_{i 1}}{\vert \vert v \vert \vert^2 - \vert \vert v \vert \vert v_1} - \frac{\vert \vert v \vert \vert^2 \delta_{i 1} \delta_{j 1}}{\vert \vert v \vert \vert^2 - \vert \vert v \vert \vert v_1}\,.
\end{align}
Note that $\vert \vert v \vert \vert$ is sourced from the Nakagami distribution $f(N/2, N \sigma^2)$ while $v_{i}$'s are sourced from the Gaussian distribution $\mathcal{N}(0,\sigma)$. 

Direct evaluation of the distribution of $M_{i j}$ is challenging, so we will resort to some approximations. The first approximation that we shall resort to is that of large$-N$. This simplifies the calculations significantly. Another fact that we shall use to our advantage is that $M$ is an involutory matrix. This means that $M$ is symmetric and $M^2 = \mathbb{I}$. This imposes the following constraint
\begin{align}
    \sum_{l}m^2_{l i} = 1 \,,~~~~ \forall \; i \; \in\,(1,N)\,,
\end{align}
which means that the norm of each row (or column) of $M$ is equal to $1$. N\"aively one may estimate $m^{2}_{l i} \sim 1/N$, assuming that the elements of $M$ are uniformly distributed. This assumption will turn out to not be correct in general, however, the estimate will not be too far off.

To proceed with the calculation, we first seek to approximate the denominator of the terms in $M_{i j}$. First, we note that in the large $N$ limit, $\vert \vert v \vert \vert \gg v_{i}$. We can then approximate the denominator by $\frac{1}{\vert \vert v \vert \vert^2}\big(1 + \frac{v_1}{\vert \vert v \vert \vert}\big)$. This is reasonable since exact calculations\footnote{Using the Nakagami distribution for $\vert \vert v \vert \vert$.} demonstrate that $v_i/\vert \vert v \vert \vert$ scales as $1/\sqrt{N}$. This simplifies the expression for $M_{i j}$ somewhat
\begin{align}
    M_{i j} \approx \delta_{i j} - \frac{v_i v_j}{\vert \vert v \vert \vert^2} + \frac{v_i \delta_{j 1}}{\vert \vert v \vert \vert}
    + \frac{v_j \delta_{i 1}}{\vert \vert v \vert \vert} - \delta_{i 1} \delta_{j 1}
    -\frac{v_i v_j v_1}{\vert \vert v \vert \vert^3} + \frac{v_i v_1 \delta_{j 1}}{\vert \vert v \vert \vert^2}+ \frac{v_j v_1 \delta_{i 1}}{\vert \vert v \vert \vert^2}
    -\frac{v_1 \delta_{i1}\delta_{j1}}{\vert \vert v \vert \vert}\,.
\end{align}
The probability distribution of terms of the form $v_i/\vert \vert v \vert \vert$ can be determined exactly 
\begin{align}
    P\left(x \equiv \frac{v_i}{\vert \vert v \vert \vert}\right) = \frac{\left(x^2+1\right)^{-\frac{N}{2}-\frac{1}{2}}}{\,_2F_1\left(\frac{1}{2},\frac{N+1}{2};\frac{3}{2};-1\right)}\,,
\end{align}
where $x \in (-1,1)$. We determine the magnitude of the various terms in the expression $M_{i j}$ by approximating them to their mean $\pm$ standard deviation. The results are listed below
\begin{align}
    &\frac{v_i v_j}{\vert \vert v \vert \vert^2} \sim \pm \frac{1}{\sqrt{(N - 4)(N - 2)}} \sim \frac{v_j v_1}{\vert \vert v \vert \vert^2}\,,~~~~~
    \frac{v_i}{\vert \vert v \vert \vert} \sim \pm \frac{1}{\sqrt{N - 2}} \sim \frac{v_1}{\vert \vert v \vert \vert} \,, ~~~~
    \frac{v_i v_j v_1}{\vert \vert v \vert \vert^3} \sim \frac{1}{\sqrt{(N - 2)(N - 4)(N - 6)}}\,,  \notag\\
    &\frac{v^2_i}{\vert \vert v \vert \vert^2} \sim \frac{1}{\sqrt{N - 2}} \pm \sqrt{\frac{2 ( N - 1)}{(N - 2)^2 ( N - 4)}} \,,~~~\frac{v^2_i v_1}{\vert \vert v \vert \vert^3} \sim \pm \frac{\sqrt{3}}{\sqrt{(N - 2)(N - 4)(N - 6)}}  \,,~~
    \frac{v^3_1}{\vert \vert v \vert \vert^3} \sim \pm \frac{\sqrt{15}}{\sqrt{(N - 2)(N - 4)(N - 6)}}\,. \notag
\end{align}
This allows us to estimate the following form of $M_{i j}$ 
\begin{align}
    M_{1 1} &\sim  \frac{1}{\sqrt{(N-2)(N-4)}}-\frac{1}{\sqrt{(N-6) (N-4) (N-2)}}+ \frac{1}{\sqrt{N-2}}\,,\\
    M_{i i} &\sim 1 -\frac{1}{\sqrt{(N-6) (N-4) (N-2)}} -\frac{1}{\sqrt{(N-2)(N-4)}}\,,\\
    M_{1 j} &\sim \frac{1}{\sqrt{N-2}}-\frac{1}{\sqrt{(N-6) (N-4) (N-2)}}\,, \\
    M_{i j \neq 1} &\sim -\frac{1}{\sqrt{(N-2)(N-4)}}-\frac{1}{\sqrt{(N-6) (N-4) (N-2)}}\,.
\end{align}
These components need not be positive, of course. We choose the overall sign of all $M_{i j}$ to be positive since we will only be dealing with even powers of $M_{i j}$. A quick check tells us that $\sum_{l}M^2_{l i} \sim 1$ to leading order, as expected. Using this, it is straightforward to see that
\begin{align}
    \sum_{l = 1}^{N}M^{4}_{l i} = \delta_{i 1}\omega(N)+ (1 - \delta_{i 1})\left(1 + \mu(N)\right)\,, ~~~~
    \sum_{l = 1}^{N}M^{2}_{l i}M^{2}_{l j} = (\delta_{i 1} + \delta_{j 1})\nu(N)  + (1 - \delta_{i 1} - \delta_{j 1})\zeta(N) \,,\label{mijsq}
\end{align}
where the functions $\omega, \mu, \nu, \zeta$ are approximated to $\mathcal{O}(N^{-4})$. These are given below.
\begin{align}
    \omega(N) &= \frac{4 N^{3/2}+N^3-12 N+16 \sqrt{N}-81}{N^4}\,, \\
    \mu(N) &= \frac{1}{N^{9/2}}\Big(N^{3/2}-5 N^{5/2}-4 N^{7/2} -4 N^3-12 N^2-28 N+65 \sqrt{N}-46\Big)\,, \\
    \nu(N) &= \frac{2 N^{3/2}+N^3-8 N+10 \sqrt{N}-48}{N^4}\,, \\
    \zeta(N) &= \frac{11 N^{3/2}+3 N^{5/2}+4 N^2+28 N+31 \sqrt{N}+150}{N^{9/2}}\,.
\end{align}
These expressions will be of interest to us since these are exactly the type of expressions that arise in \eqref{ortho1}.

\subsubsection{Orthogonal transformation revisited}
We use the estimates obtained in \eqref{mijsq} in \eqref{ortho1} to determine the variances of the distribution of $H_{i j}$ after the orthogonal transformation. Note that the matrix $M$ is the matrix $C$ that we discussed around Eqn. \eqref{ortho1}.

Then, to leading order in $N$, we estimate the following change in variance of the elements of matrix $H$ after a Householder-type orthogonal transformation
\begin{align}
    &\langle \tilde{H}^2_{i j} \rangle \sim \beta(1 + \delta_{i j}) + (\alpha - 2\beta)\times
    \Big\{\delta_{i j}\left(1 + \mu(N) + \delta_{i 1}\left(\omega(N) - \mu(N)\right)\right) +(1 - \delta_{i j})\left((\delta_{i 1} + \delta_{j 1})\left( \nu(N) - \zeta(N) \right) +\zeta(N)\right) \Big\}\,.\label{htildeij}
\end{align}
It is instructive to unpack this expression term-wise
\begin{align}
    \langle \tilde{H}^2_{1 1} \rangle \equiv A(\alpha, \beta) &\sim  2\beta + \frac{1}{N}(\alpha - 2\beta)\,, \label{fA}\\ 
    \langle \tilde{H}^2_{i i \neq 1}  \rangle \equiv B(\alpha, \beta) &\sim \alpha - \frac{4}{N}(\alpha - 2\beta)\,, \label{fB}\\
    \langle \tilde{H}^2_{1 j} \rangle \equiv C(\alpha, \beta) &\sim \beta + \frac{2}{N}(\alpha - 2\beta)\,,  \label{fC}\\
    \langle \tilde{H}^2_{i \neq j \neq 1}  \rangle \equiv D(\alpha, \beta) &\sim \beta + \frac{3}{N^2}(\alpha - 2\beta)\,,  \label{fD}
\end{align}
where we keep only the leading order term in $N$. This implies that starting with a matrix $H(\alpha, \beta)$ (where $\alpha$ and $\beta$ stand for the variances of the diagonal and off-diagonal elements), we obtain the following matrix after orthogonal transformation
\begin{align}
\begin{pmatrix}
\mathcal{N}(0,\alpha) & \mathcal{N}(0,\beta) & \dots & \mathcal{N}(0,\beta)\\ 
\mathcal{N}(0,\beta) & \mathcal{N}(0,\alpha) & \dots & \mathcal{N}(0,\beta)\\
\mathcal{N}(0,\beta) & \mathcal{N}(0,\beta) & \dots & \mathcal{N}(0,\beta) \\
\vdots & \vdots & \ddots & \mathcal{N}(0,\alpha) \\
\end{pmatrix} \rightarrow 
\begin{pmatrix}
\mathcal{N}(0,A) & \mathcal{N}(0,C) & \dots & \mathcal{N}(0,C)\\ 
\mathcal{N}(0,C) & \mathcal{N}(0,B) & \dots & \mathcal{N}(0,D)\\
\mathcal{N}(0,C) & \mathcal{N}(0,D) & \dots & \mathcal{N}(0,D) \\
\vdots & \vdots & \ddots & \mathcal{N}(0,B) \\
\end{pmatrix}\,.\label{eqABCD}
\end{align}
Note that an orthogonal Householder transformation changes the structure of the matrix. All off-diagonal elements no longer retain the same distribution, nor do all diagonal elements. 

In our calculations so far, we have made one crucial assumption: the elements of the matrix remain approximately Gaussian distributed under the orthogonal transformation. This is justified since we are working with large matrices and elements of an orthogonal matrix lie between $[-1,1]$. Furthermore, for large$-N$, we invoke the central limit theorem in the sum \eqref{hij} to argue that $\tilde{H}_{i j}$ is approximately Gaussian distributed. Therefore, on average, we do not expect the distribution to deviate significantly from a Gaussian distribution. 

\subsubsection{Tridiagonal Rosenzweig-Porter}
We will use the machinery developed in the previous sections to determine the approximate probability distribution of the tridiagonal matrix elements of the RP model. We recall that for this model $\alpha = \frac{1}{2 N}$ and $\beta = \frac{1}{4 N^{\gamma + 1}}$ in terms of the distribution \eqref{hetm}. The orthogonal transformation used for tridiagonalization proceeds as follows
\begin{align}
&\begin{pmatrix}
  1 & 0^T_{N - 1}\\ 
  0_{N - 1} & M^T_{N - 1}
\end{pmatrix}  \begin{pmatrix}
  a_0 & v^{T}\\ 
  v & H_{N - 1}
\end{pmatrix}  \begin{pmatrix}
  1 & 0^T_{N - 1}\\ 
  0_{N - 1} & M_{N - 1}
\end{pmatrix} = \begin{pmatrix}
  a_0 & \vert \vert v \vert \vert e^T_{1}\\ 
  \vert \vert v \vert \vert e_{1} & M^T H_{N - 1} M
\end{pmatrix}\,.
\end{align}
Here $a_0 \sim \mathcal{N}(0,\alpha)$ remains unchanged. The distribution of $\vert \vert v \vert \vert$ is the Nakagami distribution $V_{N - 1}(\beta)$. The important part now is the distribution of $M^T H_{N - 1} M$. From the discussion of the previous section, the resulting matrix has the distribution described in \eqref{eqABCD}. The matrix $M^T H_{N - 1} M$ can be rewritten in the following form
\begin{align}
    M^T H_{N - 1} M = \begin{pmatrix}
        a_{1} & v^{T}_1 \\
        v_1 & H_{N - 2}
    \end{pmatrix}\,.
\end{align}
Here $a_{1} \sim \mathcal{N}(0,A(\alpha,\beta))$ and $v_1 \sim \mathcal{N}(0,C(\alpha,\beta))$. The diagonal components of $H_{N - 2}$ follow the distribution $\mathcal{N}(0,B(\alpha,\beta))$ and the off-diagonal components follow $\mathcal{N}(0,D(\alpha,\beta))$.
The action of the next orthogonal matrix, which has the form \eqref{matQ}, is given as
\begin{align}
    Q^T H_{N} Q = \begin{pmatrix}
        a_0 & \vert \vert v \vert \vert & & 0^T_{N - 2} \\
        \vert \vert v \vert \vert & a_1 & \vert \vert v_1 \vert \vert & 0^T_{N - 3} \\
        0 & \vert \vert v_1 \vert \vert & a_2 & v^T_{2} \\
        0^T_{N - 3} & 0^T_{N - 3} & v_{2} & H_{N - 3}
    \end{pmatrix}\,,
\end{align}
where we have further broken down the matrix $S^T H_{N - 2} S$. The next diagonal component is given by $a_{1} \sim \mathcal{N}(0,A(\alpha,\beta))$ and the off-diagonal components are given by $\vert \vert v_1 \vert \vert \sim V_{N - 2}(C(\alpha,\beta))$.

To present this transformation more formally, we present the following atomic transformation

\begin{align}
O^{T}\begin{pmatrix}
\mathcal{N}(0,a) & \mathcal{N}(0,c) & \dots & \dots & \mathcal{N}(0,c)\\ 
\mathcal{N}(0,c) & \mathcal{N}(0,b) & \dots & \dots & \mathcal{N}(0,d)\\
\mathcal{N}(0,c) & \mathcal{N}(0,d) & \mathcal{N}(0,b) & \dots & \mathcal{N}(0,d)\\
\mathcal{N}(0,c) & \mathcal{N}(0,d) & \mathcal{N}(0,d) & \dots & \mathcal{N}(0,d)\\
\vdots & \vdots & \ddots & \vdots & \vdots\\
\mathcal{N}(0,c) & \mathcal{N}(0,d) & \dots & \dots & \mathcal{N}(0,b)\\
\end{pmatrix}O 
= \begin{pmatrix}
\mathcal{N}(0,a) & V_{L - 1}(c) & 0 & \dots & 0\\ 
V_{L - 1}(c) & \mathcal{N}(0,A(b,d)) & \mathcal{N}(0,C(b,d)) & \dots & \mathcal{N}(0,C(b,d))\\
0 & \mathcal{N}(0,C(b,d)) & \mathcal{N}(0,B(b,d)) & \dots & \mathcal{N}(0,D(b,d)) \\
0 & \mathcal{N}(0,C(b,d)) & \mathcal{N}(0,D(b,d) & \dots &\mathcal{N}(0,D(b,d)) \\
\vdots & \vdots & \vdots & \ddots &\vdots \\
0 & \mathcal{N}(0,C(b,d)) & \mathcal{N}(0,D(b,d)) & \ddots &\mathcal{N}(0,B(b,d)) \\
\end{pmatrix}\,.\label{pres}
\end{align}

Using the prescription described in \eqref{pres}, one can recursively evaluate the approximate distribution of the tridiagonal matrix elements. 

It is interesting to note that a consequence of this algorithm is that the mean of the off-diagonal elements remains larger than the standard deviation of the diagonal elements. This can be understood by first noting that $\alpha$ in equations (56)-(59) is $\mathcal{O}(1)$ for the RP model, while $\beta$ is proportional to $1/N^\gamma$. The standard deviation of the diagonal elements then scales at most as $1/N$ (without rescaling, near the beginning of the Krylov chain). It is even more strongly suppressed towards the middle of the chain. The mean of the off-diagonal elements (before rescaling) scales as $\sqrt{N} \beta$, leading to a much larger contribution. The $\sqrt{N}$ factor arises because the norm of a Gaussian random vector follows the Nakagami distribution, which has a finite mean. The diagonal elements remain Gaussian distributed.

The same algorithm can now be extended to non-Hermitian matrices by using either one choice of $M$ (which gives an Arnoldi matrix representation) or two different $M$ and $M'$ (which gives a bi-orthogonal matrix representation). The difference will arise in the distribution of the matrix elements for the two algorithms. We defer further analysis of this algorithm, with respect to its accuracy, validity, and the extent to which it captures all the moments of the Lanczos distribution, which can be of independent interest.

\section{SM7: Krylov-Eigenstate Overlap}\label{sec:s1}

A probe related to the Krylov IPR that we investigate here is the overlap of the Krylov vectors and eigenstates. For this, one can begin by considering the expansion of a Krylov basis vector in the eigenbasis. It immediately follows that the coefficient $\eta^{k}_{m}$ must follow a three-term recurrence relation for each energy $E_{m}$.
\begin{align}
    b_{n + 1}\eta^{n + 1}_{m} = (E_{m} - a_{n})\eta^{n}_{m} - b_{n}\eta^{n - 1}_{m} \;\;,\; \eta^{-1}_{m} = 0\,.\label{keigenrecA}
\end{align}
The quest is then to solve the three-term recurrence relation for an energy $E_{m}$. For the systems we consider, composed of Wigner matrices, it is known that the mean $\overline{a_{n}} = 0$. Likewise it is also known that $\overline{b_{n}}  = \sqrt{\frac{2}{\beta N}}\frac{\Gamma((1 + (N - n)\beta)/2)}{\Gamma((N-n)\beta/2)}$ \cite{Balasubramanian:2022dnj} in the ergodic phase. We shall use these facts as approximations for the coefficients in \eqref{keigenrecA}, along with the underlying assumption that there is no correlation between $b_{n}$ and $\eta^{n}_{m}$.

First, we consider the simple case of $E_{m} = 0$. This corresponds to the eigenstate roughly in the middle of the spectrum. 
\begin{align}
    b_{n + 1}\eta^{n + 1}_{m} = -b_{n}\eta^{n - 1}_{m}\,.
\end{align}
Two facts emerge immediately: $(1)$ for $E_{m} = 0$, $\eta^{n}_{m}$ is an alternating series with only even $n$ and $(2)$ for $E_{m} \neq 0$, all coefficients with odd $n$ have to be proportional to some positive power of $E_{m}$. From the expression for $b_{n}$, we find that (again, in the metallic phase) $\frac{b_{n}}{b_{n + 1}} = 1  - \frac{1}{N - n + 1}$. Therefore the recursion relation becomes
\begin{align}
    \eta^{n + 1}_{m} = -\left( 1  - \frac{1}{N - n + 1} \right)\eta^{n - 1}_{m}\,.
\end{align}
This leads to the general expression for $c^{n}_{m}$ for $n = 2k$ to be
\begin{align}
    \eta^{ 2 k + 2}_{m} = (-1)^{k}\prod_{l = 1}^{k}\left( 1  - \frac{1}{N - 2l} \right)\eta^{0}_{m}\,.
\end{align}
Thus the overlap coefficients are uniquely determined in terms of the overlap of the initial vector $\ket{\varphi_0}$ with the $E = 0$ eigenstate. This also tells us that all even Krylov vectors will have a non-zero overlap with the $E = 0$ eigenstate, provided the initial vector is not an eigenstate of the Hamiltonian.

In a general phase where the form of $b_{n}$ is not known exactly, the same conclusions will still hold. The only difference will be that the explicit form is written more formally as
\begin{align}
    \eta^{2 k + 2}_{m} = (-1)^{k}\prod_{l = 1}^{k}\left( \frac{b_{2 l + 2}}{b_{2 l + 1}} \right)\eta^{0}_{m}\,. \label{eig2overlap}
\end{align}
The situation gets more complicated for $E_{m} \neq 0$. Solutions for three-term recurrence relations have been known in the literature (see \cite{holehouse2023closed} and references therein). It is interesting to note that the general solution for these recurrence relations is functions in the Heun class.

\begin{figure}[t]
   \centering
\includegraphics[width=0.7\textwidth]{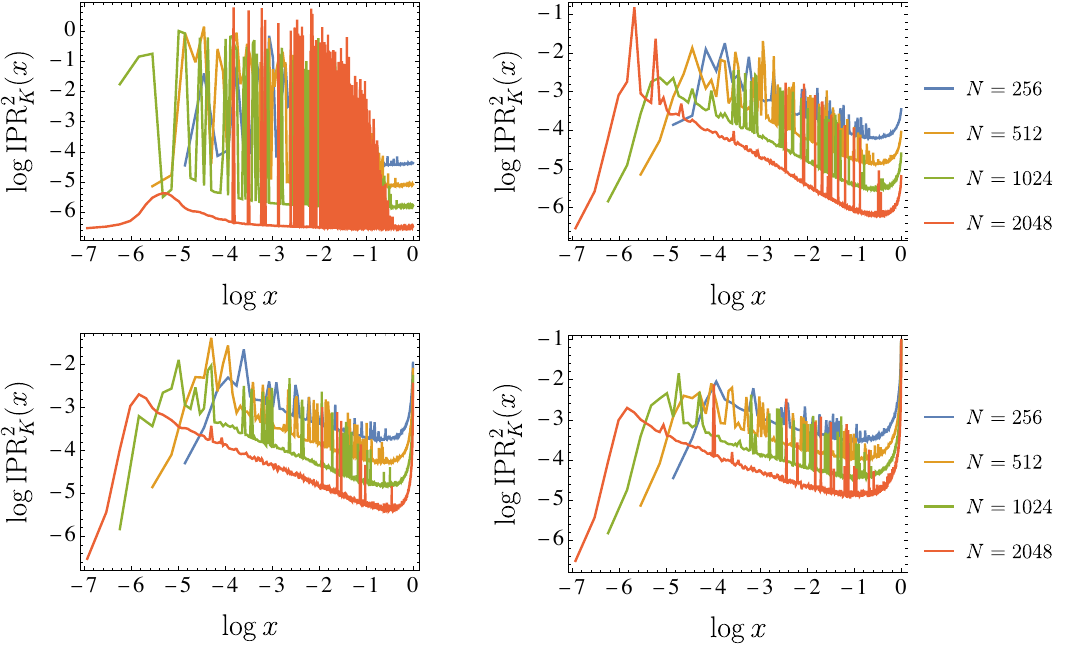}
\caption{Behaviour of $\mathrm{IPR}^{2}_{\text{K}}(\varphi_k)$ in the ergodic phase with $\gamma = 0.7$ (top, left), fractal phase with $\gamma = 1.2$ (top, right) and $\gamma = 1.7$ (bottom, left), and in the localized phase with $\gamma = 2.2$ (bottom, right). The points are taken close to the transition points $\gamma = \{1, 2\}$. } \label{fig:kipr2}
\end{figure}

We evaluate $\mathrm{IPR}^{2}_{\text{K}}(\varphi_k)$ for all $\varphi_k$ for different system sizes. Figure \ref{fig:kipr2} shows that the Krylov vectors most sensitive to $\gamma$ are the ones at the very end of the Krylov spectrum. This provides strong evidence for using $\varphi_{N - 1}$ to extract a scaling exponent. Therefore, we choose to study the scaling of the $\mathrm{IPR}^{2}_{\text{K}}$ second to last Krylov vector $\varphi_{N - 1}$.

\end{document}